\documentclass[twocolumn,showpacs,superscriptaddress,preprintnumbers,amsmath,amssymb,aps]{revtex4}
\usepackage{graphicx}
\usepackage{bm}
\usepackage{multirow}

\newcommand{\cref}[1]{(\ref{#1})}

\begin{document}
\title{Adsorption Trajectories and Free-Energy Separatrices for Colloidal Particles in Contact with a Liquid-Liquid Interface.}

\author{Joost de Graaf}%
 \email{j.degraaf1@uu.nl}
\author{Marjolein Dijkstra}%
\affiliation{%
Soft Condensed Matter, Debye Institute for Nanomaterials Science, Utrecht University, Princetonplein 5, 3584 CC Utrecht, The Netherlands
}%
\author{Ren\'e van Roij}%
 \email{r.vanroij@uu.nl}
\affiliation{%
Institute for Theoretical Physics, Utrecht
University, \\ Leuvenlaan 4, 3584 CE Utrecht, The Netherlands
}%

\date{\today}

\begin{abstract}
We apply the recently developed triangular tessellation technique as presented in [J. de Graaf \emph{et al.}, Phys.~Rev.~E~\textbf{80}, 051405 (2009)] to calculate the free energy associated with the adsorption of anisotropic colloidal particles at a flat interface. From the free-energy landscape, we analyze the adsorption process, using a simplified version of Langevin dynamics. The present result is a first step to understand the time-dependent behavior of colloids near interfaces. This study shows a wide range of adsorption trajectories, where the emphasis lies on a strong dependence of the dynamics on the orientation of the colloid at initial contact with the interface. We believe that the observed orientational dependence in our simple model can be recovered in suitable experimental systems.
\end{abstract}


\maketitle

\section{\label{sec:intro}Introduction}

The adsorption of colloidal (nano)particles at liquid-liquid interfaces is not only of scientific interest, but also relevant for industry. The formation of two-dimensional structures~\cite{pieran0,twoD}, which may be utilized in photonic bandgap materials, and the stabilization of Pickering emulsions~\cite{stapick}, are two examples of possible applications. There is also an impetus to theoretically describe the colloidal adsorption process more accurately, in order to gain a deeper insight into the mechanisms at work in experimental systems. Especially two-dimensional (2D) fluid phenomena, and phase transitions in 2D fluids of anisotropic particles~\cite{basa,vermant} are of interest. The tunability and variety of colloidal particles currently available, coupled with our still limited knowledge on colloidal adsorption, leaves the study of their interfacial phenomena an open field. 

To better understand the complex systems~\cite{strange,strange_th,selfas,right,basa,vermant} which arise when colloidal particles are brought in contact with an interface, we developed a method to determine the free energy associated to the adsorption of a single particle~\cite{paper0}. This free-energy calculation is based on similar surface tension arguments as proposed by Pieranski~\cite{pieran0}, in his ground breaking study of colloidal adsorption phenomena. The model in our investigations encompasses surface and line tension, but disregards interfacial deformation and electrostatic effects. Our contribution to this theoretical description is the triangular tessellation scheme~\cite{paper0}, which allows us to efficiently determine the adsorption free energy of an anisotropic colloid. In this paper we apply our technique to perform an initial investigation into the dynamics that occur when a colloid attaches at the interface and relaxes to its preferred configuration. 

The dynamics we study in our model is a simplified version of Langevin dynamics~\cite{lange,dhont}. From the adsorption free energy, which acts as a potential energy on the colloid, a vector field of the adsorption force is determined. This vector field is then studied by flow lines. For convenience, the anisotropy of the particle is only considered in the free energy and not in the friction tensor, which for interfacial systems would also be position dependent. Our choice allows us to examine general trends and gross features, and to showcase our numerical method, whilst at the same time give results which resemble those one would expect for actual physical systems. We consider the time scale associated with the behavior predicted by our model, and conclude that there are parameters for which the dynamics are sufficiently slow to allow these effects to be observed in experimental systems. 

In this paper we focus on three systems to point out the complexities that anisotropy can induce in the adsorption process. As a reference system we discuss the adsorption of an ellipsoidal particle for several conveniently chosen system parameters. Subsequently, we proceed to describe the same system for a cylindrical particle to show the effects of shape. The free-energy landscape for cylinders is more complicated, leading to metastable adsorption configurations with a large stability domain in the adsorption region. Finally, we study a cylindrical colloid with aspect ratio $1$. The occurrence of interfacial adsorption here is strongly dependent on the way in which the particle makes initial contact with the interface. For some initial configurations there will be no adsorption at the interface, depending on the value of the three-phase contact angle of the particle and the two liquids.

In conclusion, we apply our triangular tessellation scheme to determine the adsorption behavior of non-convex particles via flow-line dynamics. We show that for the systems considered the adsorption mechanism is far more complex than was previously believed. The effects of including interfacial deformation, and anisotropic friction, certainly merit further investigation, but go beyond the scope of our initial investigation. 

\section{\label{sec:method}Method}

\subsection{\label{sub:theoretical}Theoretical Considerations}

\begin{figure}
\includegraphics[height=3.25in]{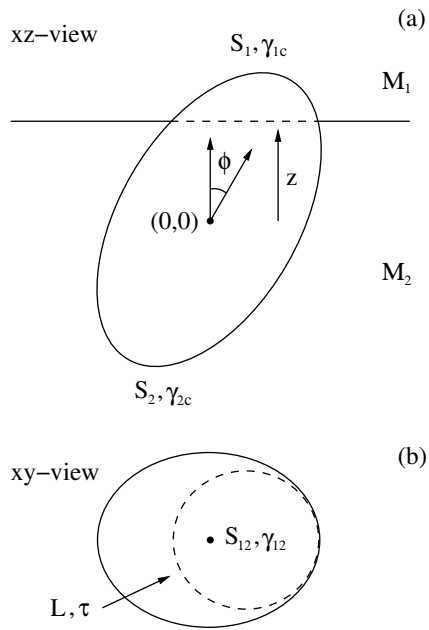}
\caption{\label{fig:config} Two representations of an ellipsoidal colloid adsorbed at a flat interface, located at depth $z$ measured from the center of the colloid $(0,0)$. The $xz$ view (a) shows the two media, their dividing interface, and the polar angle $\phi$ which the colloid's rotational symmetry axis makes with the interfacial normal. The surface area of the colloid above the interface is denoted by $S_{1}$, with $\gamma_{1c}$ the $M_{1}$-colloid surface tension, and the surface area of the colloid below the interface is denoted by $S_{2}$, with $\gamma_{2c}$ the $M_{2}$-colloid surface tension. The colloid excludes an area $S_{12}$ from the interface, the region enclosed by dashed curve in the $xy$ view (b), of which the surface tension is given by $\gamma_{12}$. The length $L$ of this dashed curve is the contact line length and the corresponding line tension is denoted by $\tau$. The solid curve in the $xy$ view indicates the colloid's outline.}
\end{figure}

We consider the adsorption of a solid uniaxial convex colloid at a planar oil-water interface separating two homogeneous half spaces of oil and water. The interfacial normal coincides with the $z$ axis of our coordinate frame, see Fig.~\ref{fig:config}. The \emph{depth} $z$ of the interface is measured with respect to the center of the particle. We locate the origin of our system at the particle's center. Medium 1 ($M_{1}$) is the half space above and medium 2 ($M_{2}$) is the half space below the interface. The \emph{polar angle} $\phi \in [0, \pi/2]$ gives the angle between the colloid's rotational symmetry axis and interfacial normal. For simplicity, capillary effects due to the presence of the colloid at the interface are neglected. A simple calculation~\cite{paper0} shows that gravity does not play a significant role in interfacial adsorption for most colloidal systems, and it is hence ignored.

The adsorption free energy of the colloid contains terms depending on three surface areas with corresponding surface tensions: (i) the surface area of the colloid above the interface $S_{1}$, (ii) the colloid's surface area below the interface $S_{2}$, and (iii) the surface area excluded from the interface by the presence of the colloid $S_{12}$. The contact line, of length $L$, where the three phases meet, also contributes to the adsorption free energy. The dependence of these quantities on $z$ and $\phi$ is implicit for notational convenience. We denote the total surface area of the colloid by $S$, therefore $S_{1}, S_{2} \in [0,S]$, under the constraint $S = S_{1} + S_{2}$.

Following Pieranksi~\cite{pieran0}, the adsorption free energy can be written as
\begin{equation}
\label{eq:free}F(z,\phi) = \gamma_{12}[(S_{1} - S)\cos\theta - S_{12}] + \tau L,
\end{equation}
where $\gamma_{12}$ is the $M_{1}$-$M_{2}$ surface tension, and $\tau$ is the line tension. This free energy, Eq.~\cref{eq:free}, is defined with respect to a reference point, namely $F = 0$ when the colloid is completely immersed in $M_{1}$. The \emph{contact angle} $\theta$ is introduced via Young's equation $\gamma_{12} \cos\theta = \gamma_{1c} - \gamma_{2c}$~\cite{young}, with $\gamma_{1c}$ and $\gamma_{2c}$ the $M_{1}$-colloid and the $M_{2}$-colloid surface tension respectively. $F(z^{*},\phi)$ is made dimensionless and scale invariant by dividing the free energy by $\gamma_{12}S$ ($\gamma_{12} \ne 0$) and writing $z = z^{*} \sqrt{a^{2}+2b^{2}}$. Here $a$ is the rotational symmetry semiaxis, $b$ the perpendicular semiaxis, and $m \equiv a/b$ the \emph{aspect ratio}. We thus obtain
\begin{equation}
\label{eq:pierpot} f(z^{*},\phi) = \frac{F(z,\phi)}{\gamma_{12}S}  = \cos\theta(r_{1}-1) - r_{12} + \tau^{*} l,
\end{equation}
where $r_{1} \equiv S_{1}/S$ and $r_{12} \equiv S_{12}/S$ are surface area ratios,
\begin{equation}
\label{eq:taustar} \tau^{*} \equiv \frac{\tau}{\gamma_{12} \sqrt{S}},
\end{equation}
is the dimensionless line tension, and $l \equiv L/\sqrt{S}$ is a dimensionless contact line length. An extended adsorption free-energy expression, which can handle nonconvex patterned colloids, is given in Ref.~\cite{paper0}.

The \emph{adsorption configuration} is the location of the free-energy minimum in Eq.~\cref{eq:pierpot} and is denoted by $(z_{\mathrm{ad}}^{*},\phi_{\mathrm{ad}})$. The adsorption free energy for this configuration is given by $f_{\mathrm{ad}} \equiv f(z_{\mathrm{ad}}^{*},\phi_{\mathrm{ad}})$. For multiple (metastable) minima, the various $(z_{\mathrm{ad}}^{*},\phi_{\mathrm{ad}})$ are labeled with a subscript $i = 1, 2, \dots$, where the deepest minimum is given the lowest index. We denote the positive value of $z^{*}$, for which the interface just touches the top of the particle by $z^{*}_{\mathrm{det}}(\phi)$ for given $\phi$; as it is also the detachment position. Note that this value is always positive, whenever we consider $-z^{*}_{\mathrm{det}}(\phi)$ the interface just touches the bottom of the particle, because of the symmetry of the problem. When $z^{*} < -z^{*}_{\mathrm{det}}(\phi)$ or $z^{*} > z^{*}_{\mathrm{det}}(\phi)$ the colloid is completely immersed in $M_{1}$ and $M_{2}$ respectively. For a given $\phi$, $f(z^{*},\phi)$ has a minimum as a function of $z^{*}$, the location of which is denoted by $z^{*}_{\min}(\phi)$ and the corresponding free energy by $f_{\min}(\phi)$. It is possible that a single equi-$\phi$-curve $f(z^{*},\phi)$ has multiple minima. This in turn leads to multiple $z_{\min}^{*}(\phi)$, which are labeled with indices as above. Analogously, $\phi_{\min}(z^{*})$ gives the location of the minimum in $f(z^{*},\phi)$ as a function of $\phi$ for a fixed $z^{*}$, the value of which we denote by $f_{\min}(z^{*})$. Here multiple minimum curves can exist as well. Often we abbreviate $z_{\mathrm{det}}^{*}(\phi)$, $\phi_{\min}(z^{*})$, and $z_{\min}^{*}(\phi)$ to $z_{\mathrm{det}}^{*}$, $\phi_{\min}$ and $z_{\min}^{*}$, respectively.

\subsection{\label{sub:numerical}Numerical Approximation Scheme}

\begin{figure}
\includegraphics[height=3.75in]{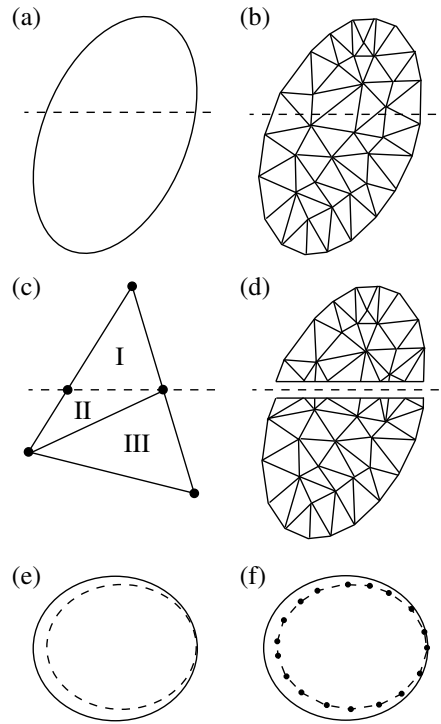}
\caption{\label{fig:numerical} A sketch of the triangular tessellation scheme in a $xz$ plane projection (a-d) and in a $xy$ plane projection (e,f). A colloid at a planar interface (a) is approximated by a polyhedron of which the faces are triangles (b). The surface of the colloid is thus effectively tessellated by triangles, which can be used to determine the approximate total surface area $\tilde{S}$. Triangles which are intersected by the interface are divided into at most three subtriangles (c). Upon this subdivision, both the area above and below the interface is approximated more accurately (d). Using the sets of triangles in graph (d), it is possible to determine $\tilde{S}_{1}$ and $\tilde{S}_{2}$, which approximate the value of $S_{1}$ and $S_{2}$ respectively. The boundary of the area cut out from the interface, the dashed line in (e), is approximated by a polygon (f). This polygon is formed by the points where the interface intersects triangles of the tessellation (b-c). The surface area $\tilde{S}_{12}$ and contact line length $\tilde{L}$ can be determined from (f).}
\end{figure}

Deriving analytic expressions for the dependence of $S_{1}$, $S_{2}$, $S_{12}$, and $L$ on $z^{*}$ and $\phi$ is highly nontrivial in general, if not impossible~\cite{paper0}. To analyze colloids adsorbed at an interface we have developed the numerical technique called \emph{triangular tessellation}, which we briefly outline here, see Fig.~\ref{fig:numerical}. Full details are given in Ref.~\cite{paper0}, including the application to more complex colloid shapes than discussed in this paper. 

Using a parametrization of the colloid at the interface, the surface and shape of the particle are approximated by a polyhedron, the faces of which are triangles, see Fig.~\ref{fig:numerical}(b). This approximation can be improved upon by reducing the overall triangle size, thus achieving better correspondence between the actual particle shape and the approximating polyhedron. The fineness of the tessellation, which is often nonequidistant, is indicated by an $n \times m$ vertex notation~\cite{paper0}. The value $nm$ is the number of vertices in the mesh, each of which is a member of 6 adjacent triangles. Modeling a 2D or 3D object in this way is a well-known computer science method and it has been successfully applied to various surface tension problems in physics~\cite{brakke0,brakke1,brakke2}. The area of each of the triangles composing the polyhedron can be determined by a simple cross product. Summing over these areas results in an approximated value for the total surface area, we denote this value by $\tilde{S}$.

When the tessellated object is intersected by a plane, some triangles of the approximating polyhedron are intersected by it, see Fig.~\ref{fig:numerical}(b). These triangles are subdivided, Fig.~\ref{fig:numerical}(c), into at most three subtriangles, which are not intersected~\cite{paper0}. After this subdivision, we determine which triangles lie above, given by the set $\tilde{\Delta}_{\uparrow}$, and below the plane, given by the set $\tilde{\Delta}_{\downarrow}$, see Fig.~\ref{fig:numerical}d. Using these sets it is possible to determine the approximated surface areas $\tilde{S}_{1} = \sum_{i} \tilde{\Delta}_{\uparrow,i}$, and $\tilde{S}_{2} = \sum_{i} \tilde{\Delta}_{\downarrow,i}$, where $\tilde{\Delta}_{\ast,i}$ is the surface area of the $i$-th triangle in the set $\tilde{\Delta}_{\ast}$. It follows that $\tilde{S}_{1} + \tilde{S}_{2} = \tilde{S}$.

From the polyhedral approximation the points are extracted where the plane intersects ribs of the triangles that tessellate the colloid, see Fig.~\ref{fig:numerical}(c,f). These are used to obtain the approximate surface area excluded from the interface by the colloid, say $\tilde{S}_{12}$, and the approximate contact line length, say $\tilde{L}$. The points of intersection form a polygon in the plane, Fig.~\ref{fig:numerical}(f), the area of which is determined using a trapezium rule type integration or a modified version of Green's integral theorem~\cite{paper0}. With either method, it is important to take steps to minimize numerical uncertainty~\cite{paper0}.

\subsection{\label{sub:traj}Adsorption Trajectories and Separatrices}

Applying the triangular tessellation technique to solve Eq.~\cref{eq:pierpot} as a function of $z^{*}$ and $\phi$ yields a free-energy landscape (on a grid). This landscape is studied in various ways, using {$z^{*}$-} or $\phi$-sections~\cite{paper0}, and by $z^{*}_{\min}(\phi)$ and $\phi_{\min}(z^{*})$ curves with respective corresponding $f_{\min}$. It is also possible to examine the negative gradient at each point of the adsorption free-energy landscape
\begin{eqnarray}
\nonumber \mathcal{F}(z^{*},\phi^{*}) & = & -\mathbf{\nabla}f^{*}(z^{*},\phi^{*}); \\ 
\label{eq:grad} & \equiv & - \left( \hat{z}^{*} \frac{\partial}{\partial z^{*}} + \hat{\phi}^{*} \frac{\partial}{\partial \phi^{*}} \right)f^{*}(z^{*},\phi^{*}) ,
\end{eqnarray}
with $\hat{z}^{*}$ and $\hat{\phi}^{*}$ unit vectors, $\pi \phi^{*} \equiv \phi$, and $f^{*}(z^{*},\phi^{*}) \equiv f(z^{*},\phi)$; thus obtaining a \emph{vector field of adsorption force} $\mathcal{F}$. Note that with this choice for $\phi^{*}$ and $f^{*}$ the domain is bounded by $(z^{*},\phi^{*}) \in [-1,1] \times [0,0.5]$. It should be emphasized that the adsorption free energy acts as a potential energy for the colloid, which can be differentiated with respect to its macroscopic coordinates, obtaining a force; the microscopic coordinates of the fluid have been integrated out to yield the tension terms. 

The vector field is studied by examining four features: (i) flow lines, (ii) minima and maxima, (iii) saddle points, and (iv) separatrices. We will show that for certain particle species there are multiple minima in the adsorption free-energy landscape, also see Ref.~\cite{paper0}. Each of these minima is surrounded by a region to which that particular minimum is attractive. That is to say, all flow lines which originate from points in this region reach that minimum when $t \rightarrow \infty$, with $t$ time. A \emph{flow line}, $\eta(t) = \left(z^{*}(t),\phi^{*}(t)\right)$, is defined to follow the path of steepest descent from its starting point at $\eta(t = 0)$. It is a solution to the differential equation
\begin{equation}
\label{eq:flowlinedyn} \dot{\eta}(t) \equiv \frac{\partial \eta(t)}{\partial t} =\mathcal{F}(\eta(t)),
\end{equation}
where the dot denotes the time derivative and the solution is fixed by imposing the initial point $\eta(t = 0)$. We refer to these flow lines as \emph{adsorption trajectories} to stress their relation with the physical path followed by a particle when adsorbing to the interface. The dividing line between two attractive regions is referred to as the \emph{separatrix}. This separatrix can contain (local) maxima and saddle points.

\subsection{\label{sub:time}Timescale of Colloid Motion}

It is desirable to review the concept of time as introduced by solving the differential equation $\dot{\eta}(t) =  \mathcal{F}(\eta(t))$. The variable $t$ in this equation is mathematically a parametrization parameter. This parameter is however related to physical time, since it is relative to the size of the force exerted at a point $\eta(t)$, but not scaled according to a friction tensor. In the Langevin equation~\cite{lange,dhont} for a sphere in a homogeneous medium, the time to travel over a short distance $\Delta l$ is equal to $M \xi \Delta l / \vert \mathbf{A} \vert$, where $M$ is the mass, $\xi^{-1}$ the Brownian time, and $\mathbf{A}$ is the force. We have a similar situation for our parametrization, save the mass and friction coefficient. The $1/\vert \mathcal{F} \vert$ is the reduced time that it takes to move along a flow line through a specific point. Note that it diverges for extrema in the free-energy landscape, which is in accordance with the particle being stationary there. We analyze the relation between the reduced time $t$ and the actual time $t_{r}$ in more detail in the ``Discussion'' section.

The flow line for a specific initial point is determined by linear differential solver which employs steepest descent. In this scheme the time parameter is determined as follows. The time in which step $i$ is taken, $\Delta t_{i}$, is defined as the ratio of the step size $\Delta l_{i}$ and the force at the initial point of the step, i.e., $\Delta t_{i} = \Delta l_{i} / \vert \mathbf{\nabla}f_{i} \vert$. We thus obtain a series of $\Delta t_{i}$. The time required to arrive at point $n$ along the flow line, say $t(n)$, from the initial point is then given by
\begin{equation}
\label{eq:time}t(n) = \sum_{i=1}^{n-1} \Delta t_{i}.
\end{equation}
In the limit of infinitesimally small step size, the solution to the vector field with the initial condition $\eta(t = 0)$ and the proper time dependence are obtained.

\section{\label{sec:resu}Results}

All data have been produced using the triangular tessellation technique~\cite{paper0} on $500^{2}$-vertex grids heterogeneously mapped to the object under consideration. To determine the landscape, the value of $f^{*}(z^{*},\phi^{*})$ is calculated on 250 by 250 nonequidistant grid points in $z^{*} \in [-z_{\mathrm{det}}^{*},z_{\mathrm{det}}^{*}]$ and $\phi^{*} \in [0,0.5]$. The data on this mesh is then interpolated with a 3rd order interpolation scheme to yield the full landscape. The accuracy of the triangular tessellation has been verified by a semianalytic method~\cite{paper0}. In this case, the technique has been proven to have a relative uncertainty of less than $10^{-4}$ per data point. The flow lines are determined by means of a linear differential solver, and contain over 1000 steps per line. The step size is reduced until subsequent reductions result in a relative difference between the curves of less than $10^{-3}$. This uncertainty is determined by summing over all points, the difference in position between two successive reductions and dividing by the total path length. The differential solver is considered to have converged on the solution of the vector field when the difference is less than $10^{-3}$. For some extreme cases we are forced to work with less precision, as we will indicate in the text. 

\subsection{\label{sub:ellipso}Adsorption of Ellipsoidal Particles}

\begin{figure}
\includegraphics[width=3.375in]{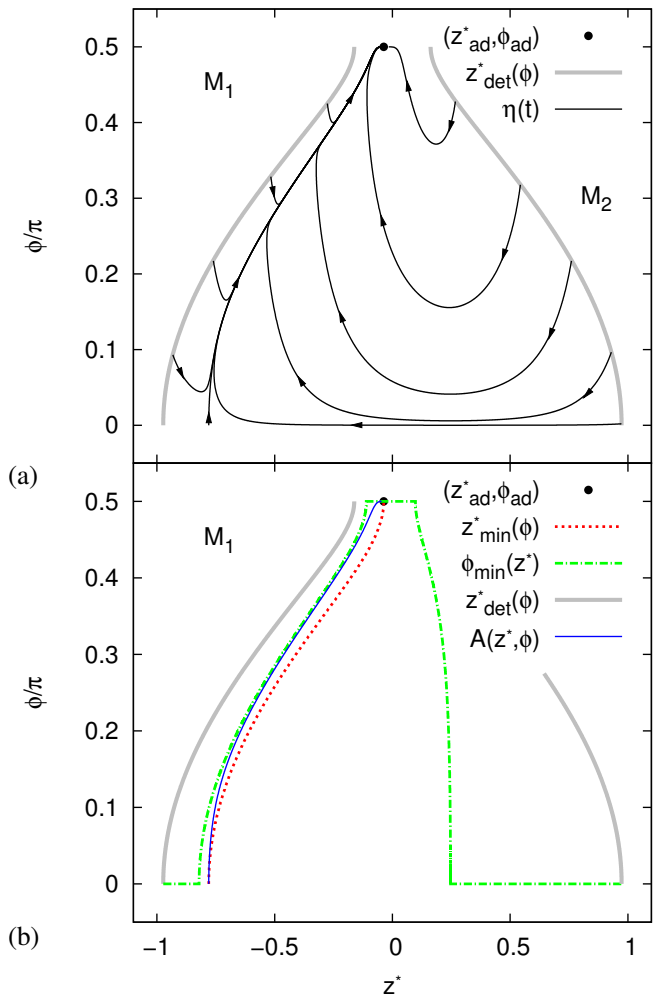}
\caption{\label{fig:ell_flow} (Color online) Graph (a) shows adsorption trajectories $\eta(t)$ for an ellipsoidal particle with aspect ratio $m = 6$ for $\cos\theta = -0.5$ and $\tau^{*} = -0.1$. The arrow heads indicate the direction of colloid motion through the free-energy landscape, the dot gives the location of the minimum, i.e., the adsorption configuration, and the thick gray lines the $\pm z^{*}_{\mathrm{det}}(\phi)$ curves. The symbols $M_{1}$ and $M_{2}$ indicate that the colloid is completely immersed in the respective media, whenever its $z^{*}$ value is to the right or left of the $\pm z^{}_{\mathrm{det}}(\phi)$ curve. Graph (b) shows the $\pm z^{*}_{\mathrm{det}}(\phi)$ curves in gray (thick, solid), the $z^{*}_{\min}(\phi)$ curve in red (thick, dots), the $\phi_{\min}(z^{*})$ curve in green (thick, dash-dot), and the attractor $A(z^{*},\phi)$ in blue (thin, solid). The location of the minimum is again given by a dot. Note that the $z^{*}_{\min}(\phi)$ and the $\phi_{\min}(z^{*})$ curves in (b) indeed intersect with points on the adsorption trajectories where the tangent is vertical and horizontal respectively, when (a) and (b) are superimposed.}
\end{figure}

Figure~\ref{fig:ell_flow}(a) shows adsorption trajectories $\eta(t)$ for an ellipsoidal particle with aspect ratio $m = 6$ for $\cos\theta = -0.5$ and $\tau^{*} = -0.1$. In Fig.~\ref{fig:ell_flow}(b) the $\pm z^{*}_{\mathrm{det}}(\phi)$, the $z^{*}_{\min}(\phi)$, the $\phi_{\min}(z^{*})$, and the attractor $A(z^{*},\phi)$ curves for this landscape are given. The location of the minimum is indicated with a dot. The term \emph{attractor} is introduced here to describe a feature in the vector field of adsorption force to which the flow lines are attracted, as can be seen in Fig.~\ref{fig:ell_flow}(a). Mathematical analysis shows that for points on the attractor one of the eigenvectors of the Hessian matrix [$(\nabla \nabla^{\mathrm{T}}) f$] is in the direction of the gradient, with a positive sign, and the other is perpendicular to it, with a negative sign. However, the gradient is nonzero for points on the attractor, which are therefore not saddle points. Note that the attractor lies between the $z^{*}_{\min}(\phi)$ and $\phi_{\min}(z^{*})$ curves, for which one of the components of the gradient vector is zero. 

By analyzing all points on the landscape, we conclude that the entire region between the $\pm z^{*}_{\mathrm{det}}(\phi)$ curves with $\phi \in [0,\pi/2]$ is attracted to the single minimum with $(z^{*}_{\mathrm{ad}},\phi_{\mathrm{ad}}) \approx (-0.0368,0.5\pi)$. This was to be expected on the basis of our observations in Refs.~\cite{paper0}. The flow lines in Fig.~\ref{fig:ell_flow}(a) give a rather abstract picture of the colloidal motion through the interface. The main point of introducing this ellipsoidal system is to have a basis for comparison, when we study cylinders in the next section.

\begin{figure}
\includegraphics[width=3.375in]{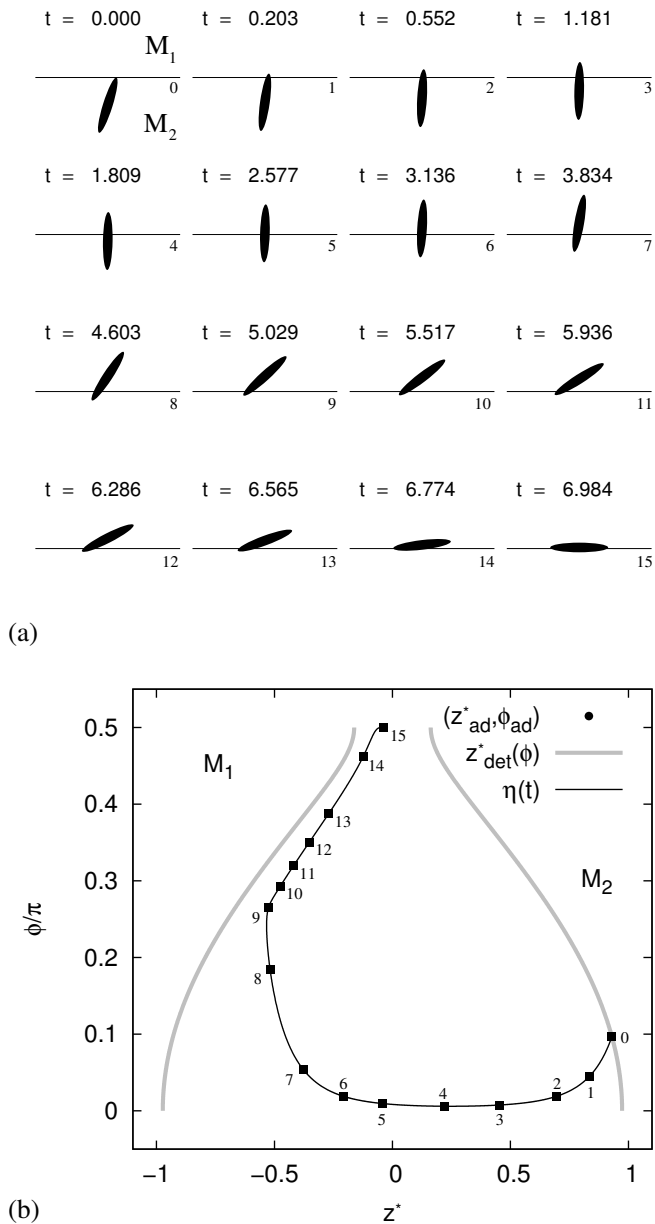}
\caption{\label{fig:ell_motion} Graph (a) shows several snapshots of the motion of an ellipsoidal colloid with $m = 6$, $\cos \theta = -0.5$, and $\tau^{*} = -0.1$ through the interface along a flow line from Fig~\ref{fig:ell_flow}. An $xz$ view of the colloid is represented by a black silhouette and the interface is indicated by a thin black line. The time, at which a snapshot is taken for each frame, is given in the top left corner. The numbers at the bottom right of the interfacial line, correspond to the numbers in graph (b). The first frame also shows the location of the two media $M_{1}$ and $M_{2}$. Graph (b) shows the $\pm z^{*}_{\mathrm{det}}(\phi)$ curves, thick gray line; the adsorption free-energy minimum, black dot (overlapped by point 15); the flow line for which the snapshots are taken, thin black curve; and the $(z^{*},\phi)$ for which the snapshots in graph (a) are taken, numbered black squares. Again the location of the media is indicated by the symbols $M_{1}$ and $M_{2}$.}
\end{figure}

To illustrate the behavior of the particle along one of the adsorption trajectories in Fig.~\ref{fig:ell_flow}(a), we have included Fig.~\ref{fig:ell_motion}, which shows the time dependent movement. Figure~\ref{fig:ell_motion}(a) shows several snapshots of the motion along the flow line with $\eta(t=0) \approx (0.930,0.0965\pi)$ in Fig~\ref{fig:ell_flow}. Figure~\ref{fig:ell_motion}(b) shows the location of the snapshots on the flow line. Here it should be stressed that the minimum is reached only at $t \rightarrow \infty$. The final snapshot shows a configuration, which is reached at a finite time, very close to the minimum.  

From Fig.~\ref{fig:ell_motion}(a) it becomes clear that the movement of the colloid through the interface, when it is close to vertical, is quite slow when compared to the rotational part of the movement before it reaches its equilibrium configuration. The final part of the movement, reaching the equilibrium position, takes infinitely long however. Slowing down at the end occurs only when the colloid is very close to its adsorption configuration. The difference in speed between the rotational and vertical motion parts of the adsorption is caused by the difference in free-energy decay between points 0 to 7 and 8 to 15 in Fig~\ref{fig:ell_motion}(b) respectively. In part it is caused by our implicit choice of the ratio between friction coefficients for the translational and rotational part of the movement. We return to this in the ``Discussion'' section.

The results presented here are for one contact angle $\cos\theta$ and one reduced line tension $\tau^{*}$ only. Further analysis and our observations in Ref.~\cite{paper0} show that qualitatively similar results are obtained for contact angles with $-1 < \cos\theta < 0$ and $\tau^{*} < 0.0$. For $\tau^{*} > 0$ the formation of adsorption barriers, as discussed in Ref.~\cite{paper0}, hinder adsorption to the interface form either medium and consequently change the adsorption free-energy landscape sufficiently to alter the adsorption trajectories qualitatively. 

Additional representations of the movement along the adsorption trajectories shown in Fig.~\ref{fig:ell_flow}(a) are given in the Appendix. The motion studied in Fig.~\ref{fig:ell_motion} will also be revisited.

\subsection{\label{sub:cylinde}Trajectories for Cylindrical Colloids}

\begin{figure}
\includegraphics[width=3.375in]{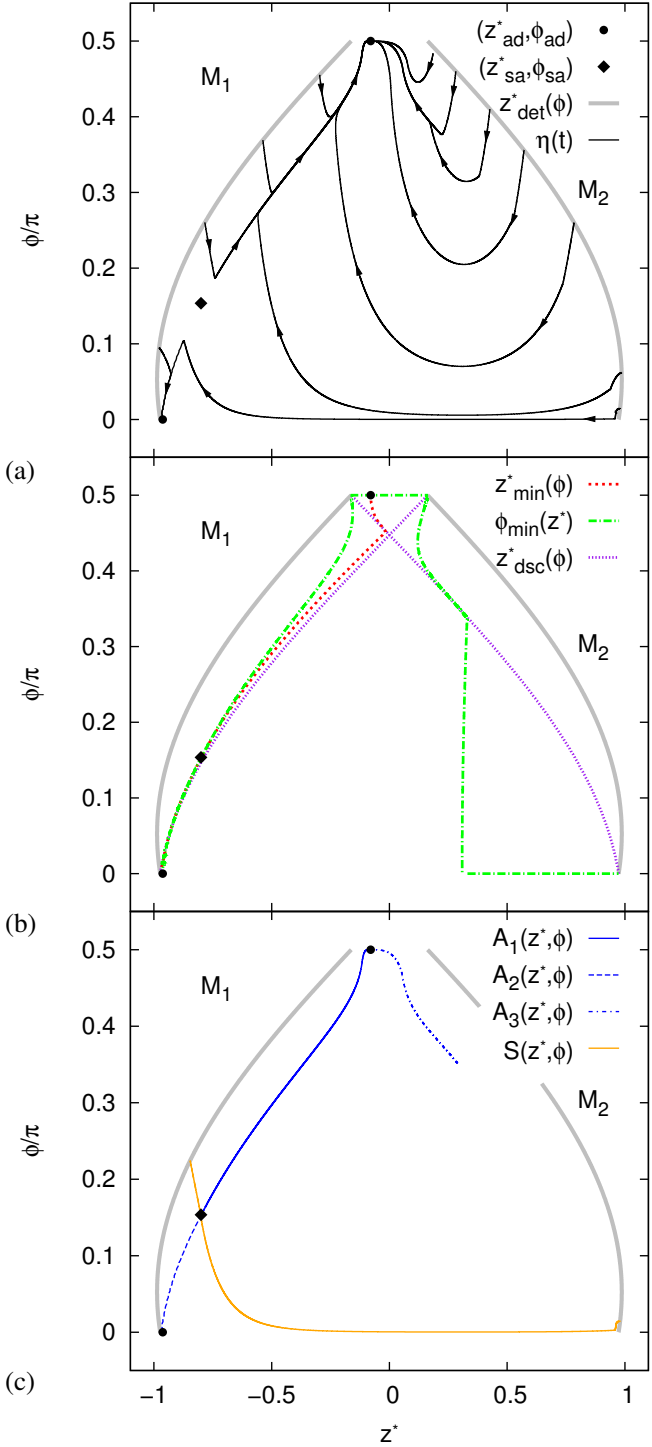}
\caption{\label{fig:cyl_flow} (Color online) Properties of the free-energy landscape for a cylinder with aspect ratio $m = 6$ for $\cos\theta = -0.5$ and $\tau^{*} = -0.1$. Graph (a) shows several adsorption trajectories $\eta(t)$ (thin black curves); the location of free-energy minima (dots), of the saddle point (diamond), and of media $M_{1}$ and $M_{2}$. Thick gray curves indicate $\pm z^{*}_{\mathrm{det}}(\phi)$. Graph (b) displays the vector field discontinuities, $z^{*}_{\mathrm{dsc}}(\phi)$, in purple (thick, dots). The $z^{*}_{\min}(\phi)$ curve is represented in red (thick, dashed) and $\phi_{\min}(z^{*})$ in green (thick, dash-dot). Graph (c) shows the attractors, $A_{i}(z^{*},\phi)$, and the separatrix, $S(z^{*},\phi)$.}
\end{figure}

Figure~\ref{fig:cyl_flow} shows the adsorption trajectories and associated curves for a cylindrical colloid with $m = 6$ when $\cos\theta = -0.5$ and $\tau^{*} = -0.1$. Despite the close similarity of these parameters to those in section A for ellipsoids, the corresponding free-energy landscape for cylinders has two minima instead of one: at $(z^{*}_{\mathrm{ad}},\phi_{\mathrm{ad}}) \approx (-0.079,0.5\pi)$ with $f_{\mathrm{ad}} \approx -0.397$ and at $(z^{*}_{\mathrm{ad}},\phi_{\mathrm{ad}}) \approx (-0.962,0.0\pi)$ with $f_{\mathrm{ad}} \approx -0.090$. For these parameters the minimum with $\phi_{\mathrm{ad}} = 0.5\pi$ is the absolute minimum, not unlike the ellipsoidal case of section A, the minimum with $\phi_{\mathrm{ad}} = 0.0\pi$ is a metastable minimum. It can be shown theoretically that the latter constitutes an configuration where one of the caps of the cylinder is flush with the interface. Thereby a surface is excluded from the interface, lowering the free energy, whilst the rest of the colloid is immersed in the preferred medium ($M_{1}$ in this case). There is only contact with $M_{2}$ on the excluded surface. This configuration can be metastable, because excluding surface area from the interface strongly lowers the free energy. 

The kinks in the adsorption trajectories of Fig.~\ref{fig:cyl_flow}(a) are directly related to ridges in the free-energy landscape caused by the sharp corners of the cylinder. These ridges are indicated by the $z^{*}_{\mathrm{dsc}}(\phi)$ curves in Fig.~\ref{fig:cyl_flow}(b), where ``dsc'' refers to the discontinuity that occurs in the gradient vector field. Note that similar to the $z^{*}_{\mathrm{det}}$ curves these are also symmetric in $z^{*}=0$, hence we will use the $\pm$ notation to distinguis between the different branches. In the bottom left corner of Fig.~\ref{fig:cyl_flow}(b) the three curves nearly coincide, but it can be shown that from left to right they are $\phi_{\min}(z^{*})$, $z^{*}_{\min}(\phi)$, and $-z^{*}_{\mathrm{dsc}}(\phi)$. From the figure it is not clearly apparent that the curve $\phi_{\min}(z^{*})$ is discontinuous at $(z^{*},\phi) \approx (0.315,0.0\pi)$. There is a small gap between the `vertical' and `horizontal' branches. The kink in the vertical branchand the starting point of the horizontal segment lie above each other. For $z^{*}_{\min}(\phi)$ the ridges in the potential landscape also induce a kink. The presence of these ridges can strongly influence the behavior of the flow lines for the cylindrical colloid's adsorption, as we will see. 

In Fig.~\ref{fig:cyl_flow}(c), we show the separatrix $S(z^{*},\phi)$ between the two minima and the three attractors $A_{i}(z^{*},\phi)$, $i = 1,2,3$, which are present in this free-energy landscape. There is a saddle point on the separatrix, where it meets with two of the attractors, $A_{1}$ and $A_{2}$. The separatrix forms the division between the regions to which the respective minima are attractive. Remarkably the size of the $(z^{*},\phi)$ domain to which the metastable minimum is attractive, is still substantial for an aspect ratio as high as $m = 6$. For a particle adsorbing to the interface from $M_{1}$, on the left-hand side of Fig.~\ref{fig:cyl_flow}(c), any colloid which touches the interface with $\phi < 0.224\pi$ will adsorb to the metastable minimum. For attachment from $M_{2}$ we find that the metastable minimum is attractive for $\phi < 0.014\pi$. Therefore, the colloid has only a very small window to adsorb to the metastable configuration from the energetically unfavorable medium. However, adsorption to the metastable configuration from the preferred medium is quite likely, because almost half of the orientations will lead to this configuration. 

The appearance of a tertiary attractor, see Fig.~\ref{fig:cyl_flow}(c), is somewhat surprising. The first and second attractor are merely the split form of a feature similar to the attractor of the ellipsoid from section A. This ``main attractor'' is split, because of the two minima, causing the separatrix to intersect it. The tertiary attractor, leading to the absolute minimum of the free-energy landscape, is caused by a subtle interplay between the right-most ridge, $z_{\mathrm{dsc}}^{*}(\phi)$, and the $\phi_{\min}(z^{*})$ curve. Note that the attractor terminates at exactly the same point as where the $\phi_{\min}(z^{*})$ curve has a kink. Also note that the presence of the tertiary attractor strongly influences the behavior of the adsorption trajectories around it, see Fig.~\ref{fig:cyl_flow}(a). 

For this cylindrical colloid with $m = 6$, we again expect qualitatively similar results for $\tau^{*} < 0$; for $\tau^{*} > 0$ adsorption barriers are found. However, since the depth of the minima is strongly dependent on the value of the contact angle $\cos\theta$, also see Ref.~\cite{paper0} for this depth when $m = 4$, the position of the separatrix will vary significantly with $\cos\theta$. We come back to this in the next section, when we discuss cylindrical particles with aspect ratio $m = 1$. 

The adsorption trajectories in Fig.~\ref{fig:cyl_flow}(a) that end up in the stable minimum, are similar to those of an ellipsoid, except around the secondary attractor, see the Appendix for examples. In Fig.~\ref{fig:cyl_motion} we consider colloid movement along one flow line which leads to the metastable minimum. Figure~\ref{fig:cyl_motion}(a) reproduces several snapshots of the motion along the flow line with $\eta(t=0) \approx (0.979,	0.014\pi)$, see also Fig~\ref{fig:cyl_flow}. Figure~\ref{fig:cyl_motion}(b) shows the location of the snapshots on the flow line. The final snapshot gives a configuration very close to the minimum, which is reached only at infinite time. It should be noted, that the motion between frames $12$ and $15$ (see Fig.~\ref{fig:cyl_motion}) could be determined only with a relative uncertainty of $\approx 10^{-2}$. This is due to the close proximity of the $\phi_{\min}(z^{*})$, $z^{*}_{\min}(\phi)$, and $-z^{*}_{\mathrm{dsc}}(\phi)$ curves, as can be seen in Fig.~\ref{fig:cyl_flow}. Consequently, the time dependence of this part of the motion has significant uncertainty, which we estimate to be at most $15$\%.

\begin{figure}
\includegraphics[width=3.375in]{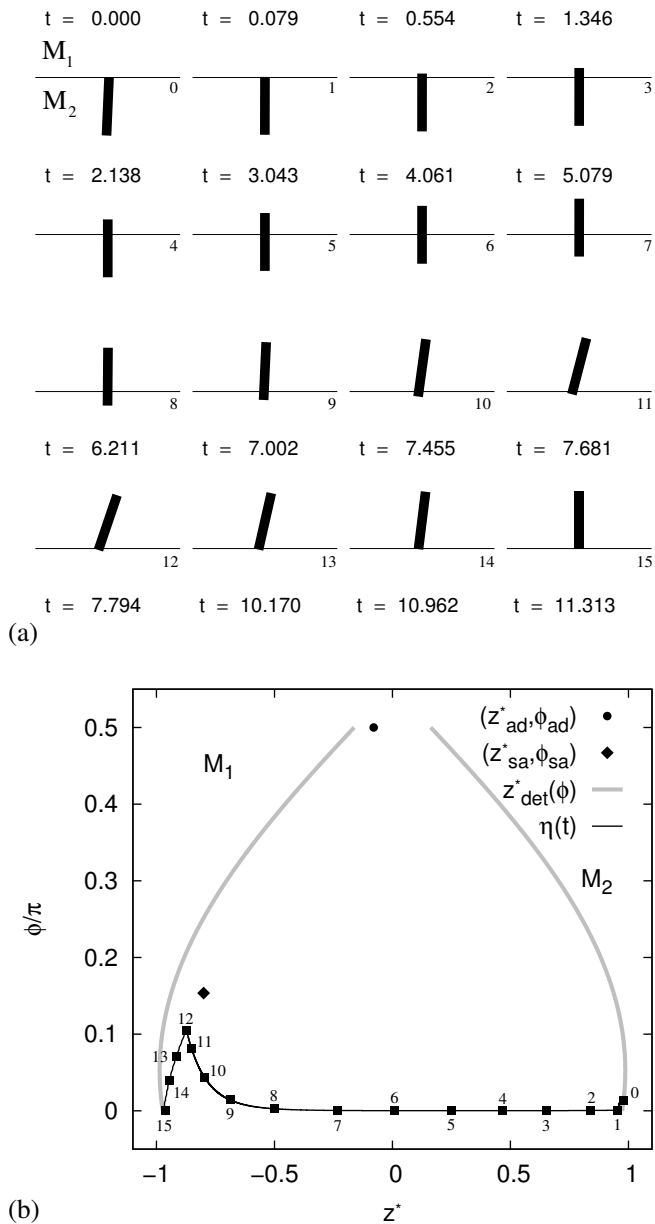}
\caption{\label{fig:cyl_motion} Graph (a) shows several snapshots of a cylindrical colloid with $m = 6$, $\cos \theta = -0.5$, and $\tau^{*} = -0.1$ adsorbing to the metastable minimum. An $xz$ view of the colloid is represented by a black silhouette and the interface is indicated by a thin black line. The time is given in the top left corner (first two rows) and the bottom left corner (last two rows). The numbers at the bottom right of the interfacial line, correspond to the numbers in graph (b). The first frame indicates the location of the two media. Graph (b) shows the $\pm z^{*}_{\mathrm{det}}(\phi)$ curves, thick gray line; the adsorption free-energy minimum, black dots (one is overlapped by point $15$); the saddle point, black diamond; the flow line for which the snapshots are taken, thin black curve; and the $(z^{*},\phi)$ for which the snapshots in graph (a) are taken, numbered black squares.}
\end{figure}

Similar to the ellipsoid in Fig.~\ref{fig:ell_motion}(a) the cylinder first moves almost vertically through the interface (frame $1 - 8$), before it tilts slightly (frame $9 - 11$). However, when it touches the interface with one point of the edge of one of the end caps, see frame $12$, it does not continue to tilt in the same direction. At this point it is energetically favorable to move into the metastable minimum by tilting back to the vertical orientation (frame $13 - 15$). Finally the colloid comes to rest with one end cap flush with the interface at $t \rightarrow \infty$. Tilting back when the edge of the end cap makes contact with the interface, frame $12$, is an indication that the discontinuity ridges in the landscape act as a dynamical ``barrier''. 

We have thus shown that if a particle can exclude a relatively large area from the interface by one of its ends, the dynamics and possible adsorption configurations are strongly influenced. Both an ellipsoid and a spherocylinder do not have a secondary minimum and therefore cannot exhibit the vertical adsorption of single particles. 

\subsection{\label{sub:separat}Separatrices and Special Configurations}

\begin{figure}
\includegraphics[width=3.375in]{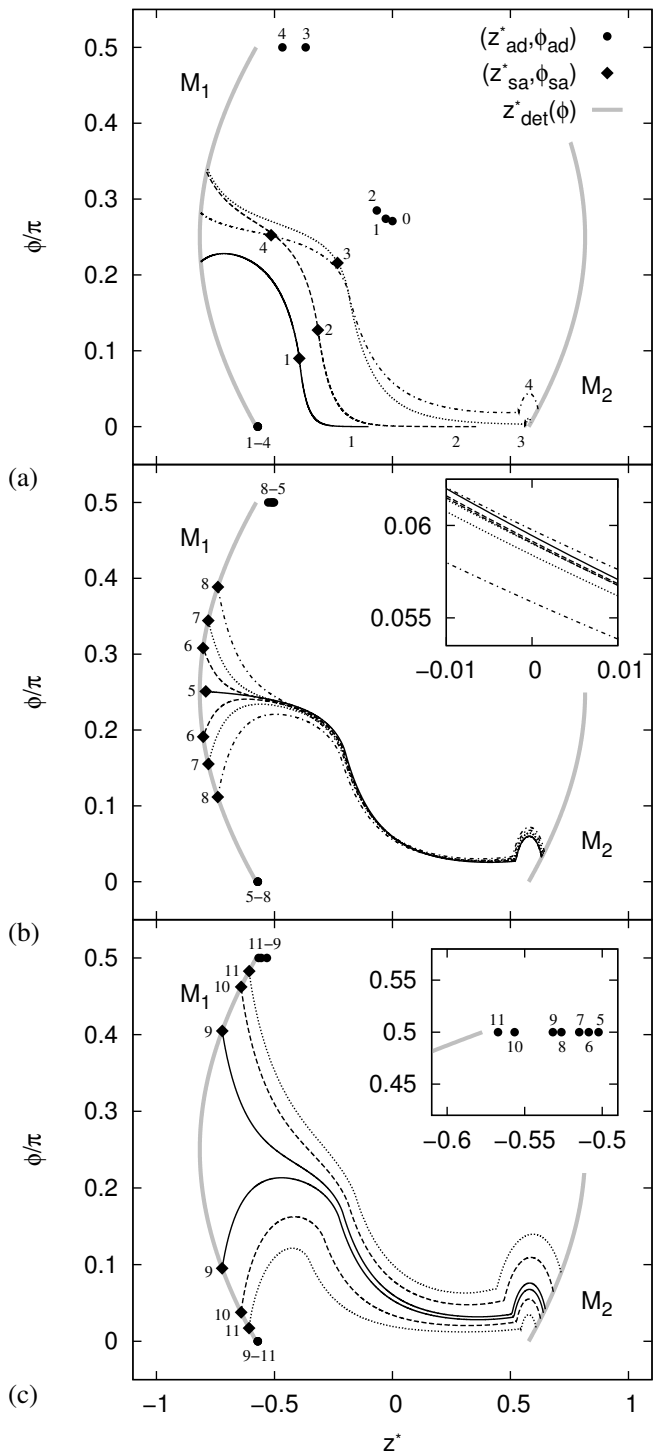}
\caption{\label{fig:cyl_sep} Properties of the free-energy landscape of a cylinder with aspect ratio $m = 1$ for $\tau^{*} = 0$ and several contact angles, labeled as follows in~(a)~$\cos\theta =$~$0.0$~$(0)$, $-0.1$~$(1)$, $-0.2$~$(2)$, $-0.4$~$(3)$, $-0.6$~$(4)$; in~(b)~$\cos\theta =$~$-0.7$~$(5)$, $-0.72$~$(6)$, $-0.74$~$(7)$, $-0.78$~$(8)$; in~(c)~$\cos\theta =$~$-0.8$~$(9)$, $-0.9$~$(10)$, $-0.95$~$(11)$. Minima are given by thick dots, saddle points by diamonds, separatrices by black lines, and the $\pm z^{*}_{\mathrm{det}}(\phi)$ curves by thick gray curves. The inset in (b) shows an enlargement of a piece of graph (b). The location of some of the minima in graphs (b) and (c) are shown in the inset in (c).}
\end{figure}

In this section we study the free-energy landscape of a short cylindrical colloid with aspect ratio $m = 1$, $\tau^{*} = 0$ and several $\cos \theta$. Figure~\ref{fig:cyl_sep}(a) shows the minima, saddle points, and separatrices for five values of the contact angle, $\cos(\theta) = 0.0$, $-0.1$, $-0.2$, $-0.4$, and $-0.6$. In Fig.~\ref{fig:cyl_sep}(b) these three properties of the free-energy landscape are given for $\cos(\theta) = -0.7$, $-0.72$, $-0.74$, and $-0.78$. The inset shows the separatrices for a small region of Fig.~\ref{fig:cyl_sep}(b). Finally, Fig.~\ref{fig:cyl_sep}(c) indicates the location of the minima, saddle points, and separatrices for $\cos(\theta) = -0.8$, $-0.9$, and $-0.95$. The inset is included to prove that the various minima in the top left corner of the graph are all separate points, as opposed to the minima in the bottom left corner of the graph, which all have the same $z^{*}$ and $\phi$ value.

For $\cos \theta = 0$ there is a single minimum at $(z^{*}_{\mathrm{ad}},\phi_{\mathrm{ad}}) \approx (0.000,0.271\pi)$ and for $\cos \theta = -1$ there is no minimum. For all other values of $\cos \theta$ one of the minima is located at $(z^{*}_{\mathrm{ad}},\phi_{\mathrm{ad}}) \approx (-0.571,0.0\pi)$, which is exactly when one of the cylinder caps is flush with the interface. The location of the other minimum changes with the value of $\cos \theta$. One of these minima is stable and the other is metastable. The minimum with $(z^{*}_{\mathrm{ad}},\phi_{\mathrm{ad}}) \approx (-0.571,0.0\pi)$ is stable when $\cos \theta \lesssim -0.23$. The labelling order of the minima with $\phi_{\mathrm{ad}} = 0.5\pi$ in the tops of Figs.~\ref{fig:cyl_sep}(b)~and~(c) indicates the way in which these appear in the graph. As already mentioned, the inset of Fig.~\ref{fig:cyl_sep}(c) shows that these minima are indeed distinct.  

With decreasing $\cos\theta$ ($\cos\theta$ tends towards $-1$) the separatrices, denoted by the black lines, move from the left lower corner towards the center of the region enclosed by the $\pm z_{\mathrm{det}}^{*}(\phi)$ curves, the \emph{adsorption region}. The location of the secondary minimum shifts closer to $(z^{*}_{\mathrm{ad}},\phi_{\mathrm{ad}}) \approx (-0.571,0.5\pi)$. For a certain value of $\cos \theta$ the separatrix splits into two pieces. This is shown in Fig.~\ref{fig:cyl_sep}(b), where the separatrices for $\cos \theta = -0.7$, $-0.72$, $-0.74$, and $-0.78$ are given (labels 5, 6, 7, and 8, respectively). Further study indicates that for $\cos \theta \approx -0.705$ there is a transition between the single and double separatrix regime. From Figs.~\ref{fig:cyl_sep}(a)~and~(b) it is apparent that for decreasing $\cos\theta$ the position of the saddle point, denoted by a diamond, moves closer to the $-z^{*}_{\mathrm{det}}(\phi)$ curve. Around $\cos \theta = -0.705$ the saddle point lies on the $-z^{*}_{\mathrm{det}}(\phi)$ curve and is degenerate. For a normal saddle point there are two attractive and two repulsive directions, whereas here there are two repulsive and only one attractive direction.

For lower values of $\cos \theta$ the saddle point transforms into two degenerate saddle points on the boundary of the adsorption region. Both of these have only one attractive and one repulsive direction. The inset in Fig.~\ref{fig:cyl_sep}(b) clearly shows that the separatrices are disjoined over the entire adsorption region. This was to be expected, since the separatrices are flow lines for a vector field. Figure~\ref{fig:cyl_sep}(c) gives a clearer picture of the distance between the two separatrices for $\cos \theta \lesssim -0.705$. To determine these curves a tolerance of $10^{-2}$ was used for the flow line convergence algorithm. When $\cos \theta \downarrow -1$ the area between the two separatrices tends to encompass the entire adsorption region. The most interesting feature of this inter-separatrix domain is that a flow line starting in it will flow towards a point on the $-z^{*}_{\mathrm{det}}(\phi)$ curve, between the two saddle points. Along such a flow line the value of $f(z^{*},\phi)$ is always positive, yet monotonically decreasing. 

\begin{figure*}
\includegraphics[width=6.75in]{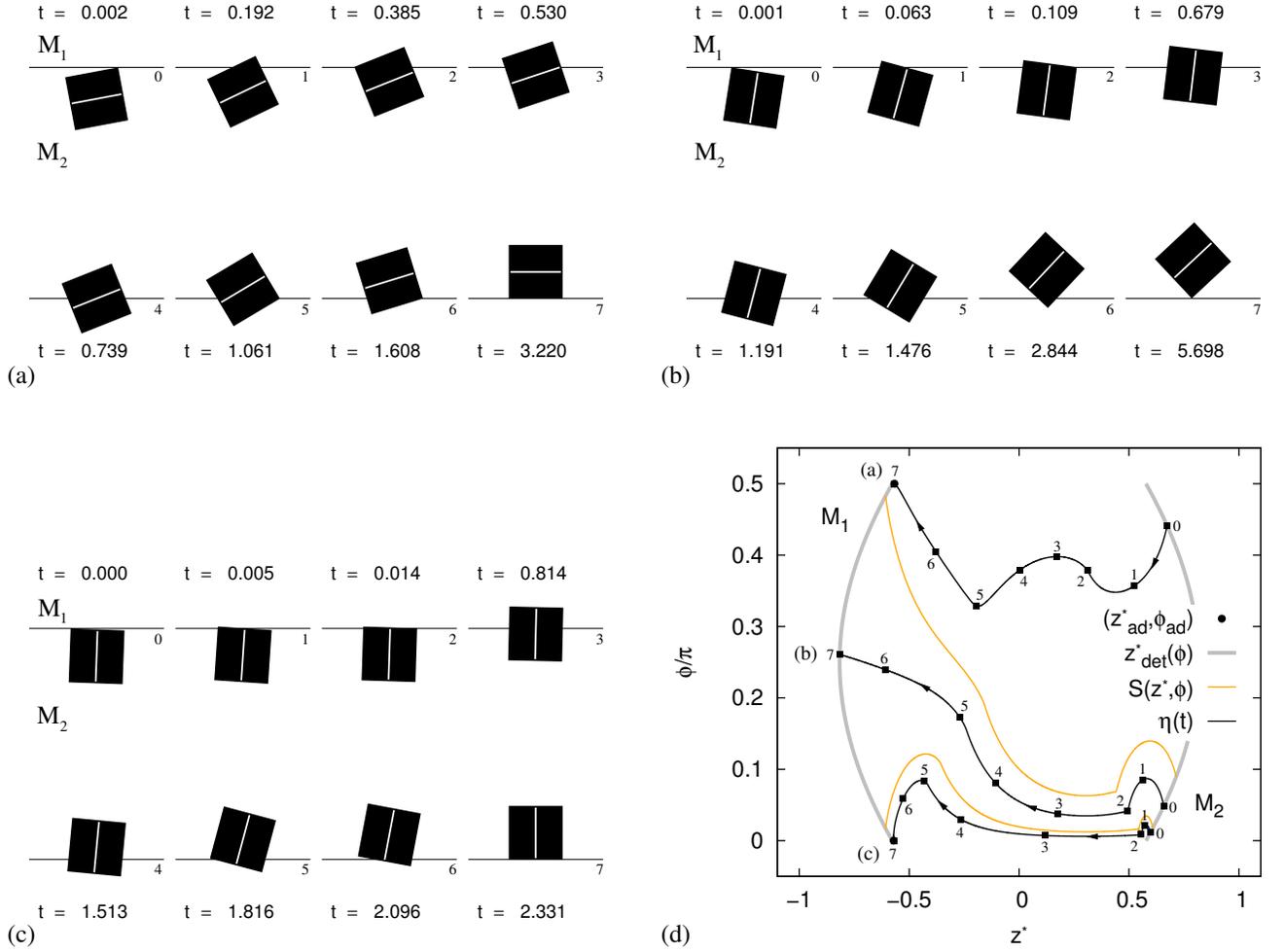}
\caption{\label{fig:cyl_mot} (Color online) Graphs (a-c) show snapshots of the motion of a cylindrical colloid with $m = 1$, $\cos \theta = -0.95$, and $\tau^{*} = 0.0$ through the interface along a flow line of the corresponding free-energy landscape. An $xz$ view of the colloid is represented by a black silhouette and the interface is indicated by a thin black line. The white line on the silhouette is used to indicate the rotational symmetry axis of the colloid. The time is given in the top left corner (first row) and the bottom left corner (second row). The numbers at the bottom right of the interfacial line, correspond to the numbers in graph (d), the graph letter is given near the ``$7$'' on each flow line. For clarity graph (d) shows the three respective flow lines $\eta(t)$ corresponding to the snapshots in graphs (a-c). The flow lines are indicated by a thin black curve, the separatrices $S(z^{*},\phi)$ [also see Fig.~\ref{fig:cyl_sep}(c)] by a thin yellow curve, the $\pm z^{*}_{\mathrm{det}}(\phi)$ by a thick gray curve, and the minima by a thick dot. In all graphs the location of the two media is given by the symbols $M_{1}$ and $M_{2}$.}
\end{figure*}

The effect of an inter-separatrix domain for certain values of the contact angle is that the adsorption of colloids can be strongly dependent on the initial configuration. The existence of adsorption minima does not necessarily imply that adsorption will take place, even if thermal fluctuations are ignored. However, the situation studied here, a cylindrical colloid with aspect ratio $m = 1$ and extreme values of the contact angle, is not representative for most colloidal systems. To illustrate the differences between the various ``adsorption'' possibilities for a cylinder with $m = 1$ and $\cos \theta = -0.95$, we have included Fig.~\ref{fig:cyl_mot}, which shows snapshots of the movement of the colloid for three different adsorption trajectories though the free-energy landscape. Figure~\ref{fig:cyl_mot}(d) gives the location of the various snapshots in Figs.\ref{fig:cyl_mot}(a)-(c). For clarity the rotational symmetry axis is given by a white line on the black silhouette of the colloid. 

The first trajectory, Fig.~\ref{fig:cyl_mot}(a), shows the colloid adsorbing to the metastable minimum at $(z^{*}_{\mathrm{ad}},\phi_{\mathrm{ad}}) \approx (-0.567,0.5\pi)$. In its final configuration, the colloid only barely penetrates the interface and its rotationally symmetry axis is perpendicular to the interfacial normal. Figure~\ref{fig:cyl_mot}(c) shows the adsorption to the primary minimum at $(z^{*}_{\mathrm{ad}},\phi_{\mathrm{ad}})=(-0.571,0.0\pi)$. Here the final configuration is when one of the cylinder caps is flush with the interface, as discussed earlier. Note that the rotational symmetry axis is indeed parallel to the interfacial normal. The intermediate series of snapshots, Fig.~\ref{fig:cyl_mot}(b), shows the colloid along a flow line in the inter-separatrix region. The colloid moves through the interface and slows down when it approaches the $-z_{\mathrm{det}}^{*}(\phi)$ curve, since the gradient tends to zero here. The final configuration included shows the cylinder with only a single attachment point on the interface, it is in essence detached. The three trajectories shown in Fig.~\ref{fig:cyl_mot} have also been included in the movie files referred to in Appendix.

\section{\label{sec:disc}Discussion}

In the previous we have seen that the dynamics of colloid motion through the interface can be very rich. In this section we analyze the relation between the proposed flow-line dynamics and the dynamics associated to experimental systems. We also discuss the elements present in real systems, which we have not incorporated in our simple model and consider how they might affect the obtained results.

\subsection{\label{sub:realtime}Estimate of the Adsorption Time}

For a dispersed particle undergoing a force, the dynamics are governed by a solution to the complete Langevin Equation~\cite{lange,dhont}. When we neglect inertia, the random force term, and limit ourselves to studying the $z$ and $\phi$ components, the equations of motion (EOMs) become
\begin{eqnarray}
\label{eq:real_z_eom} \lambda(a) \frac{\partial z(t_{r}) }{\partial t_{r}} & = & - \frac{\partial}{\partial z} F(z(t_{r}),\phi(t_{r})); \\
\label{eq:real_p_eom} \mu(a) \frac{\partial \phi(t_{r}) }{\partial t_{r}} & = & - \frac{\partial}{\partial \phi} F(z(t_{r}),\phi(t_{r})),
\end{eqnarray}
where $t_{r}$ is the ``real'' unreduced time, $F$ is our unscaled adsorption free energy, and $a$ is the long semiaxis of the particle. The prefactors $\lambda$ and $\mu$ represent the translational and rotational friction coefficients respectively. These coefficients should account for the size $a$ and the shape of the particle, as well as, the positional $z$ and orientational $\phi$ dependencies of the system. This includes the difference in viscosity between the two media, which influences the friction force, the way in which friction changes in the $z$ direction according to the colloid's $\phi$, and vice versa. We have not included these dependencies in the notation used here, because of the simplifications we apply to Eqs.~\cref{eq:real_z_eom} and~\cref{eq:real_p_eom} in the following.

Let us assume that $\lambda$ and $\mu$ are independent of $z$ and $\phi$; therefore, only changing with scaling and particle type; we return to the validity of this later. It can then be easily shown that the EOMs of Eqs.~\cref{eq:real_z_eom} and~\cref{eq:real_p_eom} allow us to regain the flow-line dynamics of Eq.~\cref{eq:flowlinedyn}, $\dot{\eta}(t) = \mathcal{F}(\eta(t))$. We require that $F(z,\phi) = \gamma_{12}S f^{*}(z^{*},\phi^{*})$, $z = \sqrt{a^{2}+2b^{2}}z^{*}$, $\phi = \pi \phi^{*}$, $t_{r} = \kappa t$, where $\kappa$ is a time scale based on the system parameters, and
\begin{eqnarray}
\label{eq:rel_fric} \mu(a) & = & \frac{a^{2}+2b^{2}}{\pi^{2}} \lambda(a) ; \\
\label{eq:unittime} \kappa & = & \frac{a^{2}+2b^{2}}{\gamma_{12} S} \lambda(a),
\end{eqnarray}
where the friction coefficients are now coupled by imposing the flow-line criterion.

For a sphere we set $\lambda(a) = 6\pi \eta_{0} a$, with $\eta_{0}$ the viscosity of the densest medium. In this way we regain the expected translational friction~\cite{einst} when the particle is free to move in that medium. Note that for this choice, we do not recover the expected rotational friction coefficient $8 \pi \eta_{0} a^{3}$~\cite{debye,mccon} in front of the time derivative $\partial \phi/\partial t_{r}$. Instead we find $18\eta_{0} a^{3}/\pi$, which differs by a factor of $9/4\pi^{2} \approx 0.23$. In retrospect, it would have been more appropriate to rescale the system such that the flow-line dynamics give the expected results for a sphere. However, we investigated the adsorption trajectories with this constraint and found qualitatively similar results.

Before discussing the accuracy of the simplifications used in our research, let us gauge the time $t_{r}$ required for a colloid to attach to the interface and relax to its final position. We assume $\lambda(a) = 6\pi \eta_{0} a$, with $\eta_{0}$ the largest of the two viscosities. For the movements studied in Figs.~\ref{fig:ell_motion}, \ref{fig:cyl_motion}, and \ref{fig:cyl_mot} the duration of the adsorption process (the part in which significant changes occur) is of the order $t \approx 5$. According to Eq.~\cref{eq:unittime}, the physical time is estimated at 
\begin{equation}
\label{eq:realtimespan} t_{r} \approx 6\pi \eta_{0} a \frac{a^{2}+2b^{2}}{\gamma_{12} S} t.
\end{equation}
Upon using this approximation we arrive at the following result, see Table~\ref{tab:time}. Here we have considered an ellipse with aspect ratios $m = 1$ and $m = 6$, sizes $a = 125$ nm, $a = 500$ nm, and $a = 2.5$ $\mu$m, which are typical colloidal length scales. For the viscosity we use $\eta_{0} = 1.5$ Pa\,s (glycerol), $\eta_{0} = 1.56 \cdot 10^{-2}$ Pa\,s (cyclohexylchloride), and $1.0\cdot10^{-3}$ Pa\,s (H$_{2}$O)~\cite{visc}, which are commonly used solvents. Surface tension values between two typical liquids are of the order $\gamma_{12} = 10^{-2}$ Nm$^{-1}$~\cite{israel} or lower than $\gamma_{12} = 10^{-5}$ Nm$^{-1}$ with the addition of surfactants~\cite{lowgam1,lowgam2}. 

\begin{table*}
\caption{\label{tab:time} The physical time $t_{r}$ required to complete the motion of the colloid through the interface on the basis of our model. We assume $t \approx 5$ and study two aspect ratios $m = 1$ and $m = 6$ for ellipsoids. Several values of the viscosity $\eta_{0}$ are considered.}
\begin{ruledtabular}
\begin{tabular}{cc|cccc}
$\gamma_{12}$ & $\eta_{0}$ & $m=1$, $a = 125$ nm & $m = 1$, $a = 0.5$ $\mu$m & $m = 6$, $a = 0.5$ $\mu$m & $m = 6$, $a = 2.5$ $\mu$m \\ 
(Nm$^{-1}$)  & (Pa\,s) &  $t_{r}$~(ms) & $t_{r}$~(ms) & $t_{r}$~(ms) & $t_{r}$~(ms) \\
\hline
\multirow{3}{*}{$10^{-2}$} & $1.5$ & $4\cdot10^{-1}$ & $2\cdot10^{0}$ & $4\cdot10^{0}$ & $2\cdot10^{1}$ \\
 & $1.56\cdot10^{-2}$ & $4\cdot10^{-3}$ & $2\cdot10^{-2}$ & $5\cdot10^{-2}$ & $2\cdot10^{-1}$ \\
 & $1.0\cdot10^{-3}$ & $3\cdot10^{-4}$ & $1\cdot10^{-3}$ & $3\cdot10^{-3}$ & $1\cdot10^{-2}$ \\
\hline
\multirow{3}{*}{$10^{-5}$} & $1.5$ & $4\cdot10^{2}$ & $2\cdot10^{3}$ & $4\cdot10^{3}$ & $2\cdot10^{4}$ \\
 & $1.56\cdot10^{-2}$ & $4\cdot10^{0}$ & $2\cdot10^{1}$ & $5\cdot10^{2}$ & $2\cdot10^{2}$ \\
 & $1.0\cdot10^{-3}$ & $3\cdot10^{-1}$ & $1\cdot10^{0}$ & $3\cdot10^{0}$ & $1\cdot10^{1}$
\end{tabular}
\end{ruledtabular}
\end{table*}

Note that it follows from Table~\ref{tab:time} that length of the adsorption process can vary significantly with the choice of system parameters. For values of the interfacial tension which are in the order $10^{2}$ Nm$^{-1}$ the adsorption process typically takes tens to hundreds of microseconds. However, upon lowering the interfacial tension, e.g., by adding surfactants, the process slows down significantly, in some cases taking upwards of a second. This is an exciting prospect, since for many systems the time scale is thus accessible to experimental techniques. It can also be shown that the particular shape of the particle, e.g, ellipsoid, cylinder, or spherocylinder, does not influence these results significantly, since $(a^{2}+2b^{2})/S$ is virtually the same for these particle types.

The limiting factor in an experiment to determine the adsorption behavior is the maximum operating frequency of the camera and not the optical elements used for imaging. For modern cameras operating frequencies of $10.000$ Hz or more are obtainable~\cite{camera}. However, we should bare in mind that our time $t_{r}$ is based on a rather crude estimate and that there can be a substantial deviation from the prediction from this value for an actual experimental system. Nevertheless, a discrepancy of a decade or two will still place many of the above mentioned systems within the experimentally observable range.  

\subsection{\label{sub:limit}Limitations of the Model and Experimental Considerations}

There are several key points of criticism which can be identified when trying to compare the dynamics predicted by our flow lines in the reduced system to real-world dynamics. These points should be taken into account in follow-up studies, but go beyond the scope of an initial investigation.

Neglecting the random thermal force component is an acceptable simplification, if we only want to study the average movement, i.e., the trajectory averaged over many adsorption events. Some care should be taken though, since it is clear from, for instance, Fig.~\ref{fig:cyl_sep}(b) that adsorption trajectories which have almost identical starting points can diverge from each other quite rapidly. This may prove problematic in establishing an average for an experimental system, especially when the magnitude of the free-energy landscape features becomes in the order of a k$_{B}$T. Neglecting the inertia is also acceptable, since most colloidal systems are strongly overdamped. Even under high potential differences $-$ a 1 $\mu$m colloidal particle can typically experience an interfacial adsorption potential in the order of $10^4$-k$_{B}$T~\cite{pieran0} $-$ colloids cannot be easily forced into the ballistic regime of motion.

Another point of concern is neglecting anisotropic effects in friction force. As we have seen in the previous, the friction coefficient in part imposes the time scale on the system. For anisotropic particles, which have a friction coefficient tensor, this time scale changes according to the direction of motion and the position of the particle. However, the difference in friction coefficient between motion along the long semiaxis of an ellipsoid and perpendicular to this semiaxis has been shown to be no greater than a factor of two~\cite{fric1,fric2,fric3,dhont}. For the rotational friction, the change is likely to be more substantial than a factor of two for strongly oblong particles. Suppose we assume proportionality with the rotational radius to the third power, as is the case for a sphere~\cite{mccon}. Then, depending on the axis of rotation, the friction coefficient varies between $\mu \propto a^{3}$ and $\mu \propto b^{3}$. Even if the proportionality is less extreme than in our simple estimate, the differences can easily be very significant. 

The difference in friction between the two media bordering the interface, can be several orders of magnitude, see for instance Ref.~\cite{visc} for the viscosity values of commonly used solvents. Such a difference in friction coefficient will play a role in describing a particular system accurately. We assumed that the timescale of the dynamics are dominated by the medium with the highest friction coefficient, or, equivalently, viscosity. This assumption holds whenever there is substantial penetration of the colloid in that medium. A more accurate approach requires us to weight the friction coefficient by the surface areas of the particle which are in the respective media. Strong differences in viscosity of the respective media most likely induce a resistance for the colloid to move into the more viscous medium. This can substantially alter the qualitative behavior of the flow lines.

For experimental systems the solvent is sometimes density-matched to the particle to eliminate the effects of gravity. Matching is however more complicated to achieve for a three component system, such as the particle near a liquid-liquid interface. To account for gravity in our model, whenever, we require the buoyancy mass of the particle. This mass can be determined by volume integration over the parts which penetrate the respective media, assuming a homogeneous density distribution. Such volumetric integrations can be performed, using a 3D analogy of our 2D surface integration scheme based on triangular tessellation. That is to say, some kind of spatial tessellation using polyhedra. Nevertheless, we have shown in Ref.~\cite{paper0} that for colloidal systems the effects of gravity will be negligible, because of the low Bond number for such particle sizes.

In many physical system, the particle and the interface are often charged by self dissociation of the surface molecules on the colloid and of those which compose the media~\cite{leuniss}. For a like-charged interface there can be a substantial charge repulsion barrier which needs to be overcome before adsorption can take place, if this barrier can be crossed at all. Including electrostatic effects into the model will substantially change the appearance of the free-energy landscape. Finally, we should stress that including interfacial deformation~\cite{oettel1,scriven} into our model should qualitatively change the results, whenever the capillary effects are dominant. Together with electrostatic effects, the results of deformation cannot be predicted at this time.  

In conclusion, we have identified the time which we expect the adsorption process to take for several experimentally feasible systems, based on our model. The results are encouraging as this time regime seems to be accessible to current experimental techniques, even though there are still quite a few caveats to be considered. 

\section{\label{sec:conc}Concluding Remarks}

We have employed the triangular tessellation technique introduced in Ref.~\cite{paper0} to determine the adsorption free-energy landscape of anisotropic colloids at a flat interface. This free energy is composed of surface and line tension contributions, but does not take into account interfacial deformation and electrostatics. We analyzed the obtained free-energy landscapes by means of a vector field of adsorption force and its associated flow lines. These flow lines are calculated using a linear steepest descent method and they are parametrized in units of reduced time. The link between these flow line dynamics and Langevin dynamics and the validity of our approximation are examined. For typical viscosities and surface tensions we predict that timescales to complete the adsorption process are in the microsecond to second regime. Such timescales are accessible for observation in modern experimental set-ups. 

Within our model, we re-establish that there is a strong dependence of the adsorption free-energy landscape on the shape of the colloidal particle. This in turn leads to a wide range of adsorption phenomena. For ellipsoidal particles we find a single equilibrium adsorption configuration with relatively simple dynamics. However, for cylindrical particles there can be two adsorption configurations, depending on the value of the contact angle. The presence of two minima in the free-energy landscape leads to a separatrix, which forms the divide between the regions to which the respective minima are attractive. The metastable minimum can have a large domain in the adsorption region to which it is attractive. This is significant, since it suggests that adsorption in unexpected configurations should be easier than previously believed. The implications of this large domain size on the experimentally observed buckling and flipping transitions~\cite{basa,vermant} merits further investigation.

We have also shown that for short cylinders the free-energy landscape allows for another type of colloid interaction with the interface, other than adsorption. Depending on the value of the contact angle, there can be a domain in the adsorption region, for which colloids do not attach to the interface, but simply move through it unhindered. This is of particular interest, since it shows, within the confines of our model, that the presence of stable and metastable adsorption configurations is not a sufficient criterion to guarantee particle adsorption. To the best of our knowledge, this is the first time that this phenomenon has been observed theoretically. 

Our simple model to calculate the free energy does not include interfacial deformation due to capillary, electrostatic, or gravitational forces; nor does it take into account inertia, anisotropy in the friction tensor, and the Brownian random force to determine the colloid dynamics. We are however optimistic that the phenomenology described here can be recovered in a more elaborate model. The method presented here thus forms the first stepping stone towards obtaining better correspondence between experiments and their theoretical description, and thus brings us to a closer understanding of colloidal adsorption phenomena.

\section{\label{sec:ackn}Acknowledgments}

M.D. acknowledges financial support by a ``Nederlandse Organisatie voor Wetenschappelijk Onderzoek'' (NWO) Vici Grant, and R.v.R. by the Utrecht University High Potential Programme.

\appendix
\section{\label{appsec:movies}Colloid Motion Though the Interface}

In this Appendix we describe the content of the EPAPS files~\cite{EPAPS} for this paper. The EPAPS files give movies for the time dependent movement along a few adsorption trajectories representative of the colloid's motion through the free energy landscapes of Figs.~\ref{fig:ell_flow}, \ref{fig:cyl_flow} and \ref{fig:cyl_mot}.

The data repository contains three folders labeled as follows: the particle type appears first, followed by the aspect ratio $m$, followed by the value of the contact angle $\cos \theta$, and finally the line tension value $\tau^{*}$ is given. The lower case ``m'' appearing in front of the numbers, is to indicate that it is a negative value, where the ``m'' stands for minus. The individual movie files are labeled similarly: an abbreviation of the type, followed by the value of the aspect ratio, and finally the coordinates of $\eta(t = 0)$ in the order $z^{*}(t=0)$, $\phi^{*}(t=0)$. The file where the value of the aspect ratio is preceded by ``A'' give the motion of the colloid along the attractor. 

Each movie file consists of a panel which shows two frames side by side. The left frame gives the $\pm z_{\mathrm{det}}^{*}$ lines in gray as well as the flow line in black. The final configuration is indicated by a black dot and the configuration at time $t$ by a red dot. The time appears in the top left corner of the right frame. A thin line divides this frame in two and serves to indicate the location of the interface. A silhouette similar to those used in Figs.~\ref{fig:ell_motion}(a),~\ref{fig:cyl_motion}(a)~and~\ref{fig:cyl_mot}(a-c) gives the $xz$ projection of the colloid. Again a white line is added to the silhouette of the cylindrical colloid with $m = 1$ to show the colloids orientation.

A few of the movie files show slight jittering of colloid along its adsorption trajectory. Especially flow lines to the metastable minimum of a cylinder with $m=6$, also see the main text. These adsorption trajectories suffered from slightly higher levels of numerical uncertainty than the others, since many features in the landscape are close together in this case.

\end{document}